\documentclass[twocolumn,floatfix,prb,showpacs]{revtex4}
\usepackage{graphicx}
\usepackage{amsmath}
\usepackage{bm}
\usepackage{hyperref}
\usepackage{soul}
\usepackage{ulem}
\usepackage[dvips]{color}

\begin{document}
    \title{Relativistic model for electron-hole pairing in the superconducting state of graphene-based materials}

\author{Mir Vahid Hosseini and Malek Zareyan}

\affiliation{Department of Physics, Institute for Advanced Studies
in Basic Sciences (IASBS), Zanjan 45137-66731, Iran}

\begin{abstract}
We propose a graphene-based model for realizing a new type of gapless condensate by pairing of electron-like (n) carriers of a Dirac cone conduction band with hole-like (p) carriers of a Dirac valance band. Ferromagnetic superconductivity (FS) in monolayer graphene or pairing between oppositely (n and p) doped layers of a double layer graphene allow for the formation of this p-n superconductivity. For FS in graphene, the p-n condensate dominates the zero temperature phase diagram at low levels of doping and high exchange fields. We show that p-n pairing with p+ip-wave symmetry presents a stable condensate phase, which can cover the phase diagram up to surprisingly strong exchange fields. Our study reveals that the characteristics of relativistic quantum physics affect the interplay between ferromagnetic ordering and superconductivity in a fundamental way.
\end{abstract}

\pacs{74.78.-w, 12.38.-t, 75.75.-c, 74.70.Wz}
\maketitle

The exclusive phenomena of superconductivity and ferromagnetism are two of the fundamental macroscopic manifestations of quantum physics governing the state of electrons in a conducting material. While the former is based on the binding of electrons with opposite spin into the so-called Cooper pairs, the later tends to align spin of electrons in order to suppress Coulomb repulsion through Pauli exclusion. The old and long-standing~\cite{FuldeFerrell,LarkinOvchin} interest in their interplay has been greatly intensified during the past two decades. The motivation has come from the experimental realization of artificial nano-scaled hybrid structures of ferromagnets and superconductors ~\cite{FSrev}, which allow for a controlled study of interplay between these competing correlations, and also by the discovery of new magnetic superconducting materials, iron-based superconductors~\cite{Irons} being the most recent ones that exhibit high transition temperatures and exciting properties.

\begin{figure}[!tb]
\begin{center}
\includegraphics[width=8.cm]{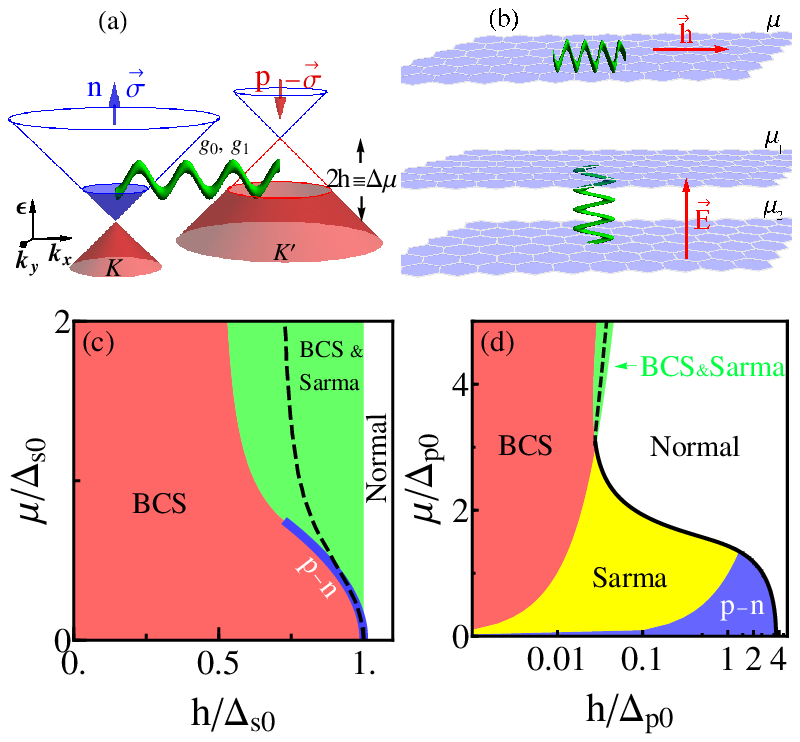}
\caption{(color online) (a) Schematic of p-n pairing (illustrated by wavy lines) between electron-like carriers of spin $\sigma$ in the Dirac cone conduction band of the $K$-valley and hole-like carriers of opposite spin in the valance band of the $K^{\prime}$-valley. The pairing can be realized in a ferromagnetic monolayer graphene (b, upper panel) with an exchange field $h$ exceeding its chemical potential $\mu$, or between layers of a double layer graphene (b, lower panel) with a difference in the chemical potentials $\Delta\mu$ (can be controlled by a perpendicular electric field $\vec E$ ) of the two layers exceeding their mean value $(\mu_1+\mu_2)/2$. The zero-temperature phase diagram of graphene-based ferromagnetic superconductivity showing dependence on $h$ and $\mu$ for s-wave (c) and p+ip-wave (d) pairing symmetry. The superconducting region consists of three phases of BCS, Sarma and p-n pairing. The dashed line in (c) and (d) indicates the instability boundary for BCS phase separating stable BCS states (left side) with the unstable (right side) ones. The S to N transition on these lines is first order. While the solid line in (d) indicates a second order S-N transition.
\label{fig1}}
\end{center}
\end{figure}

Until now, most of the studies on ferromagnetic superconductivity (FS) focus on Bardeen-Cooper-Schrieffer (BCS) limit ~\cite{Chandra,Izuyama,Dias} in ordinary electronic material with a high chemical potential, in which the electrons, as non-relativistic particles, occupy essentially parabolic energy bands having an almost constant density of states in the scale of the superconducting pairing gap. In this respect, the question of FS for relativistic fermions has remained unanswered.  In addition to its emergence in the context of high temperature and heavy fermion superconductivity ~\cite{RelSuper}, color superconductivity in dense quark matter at low temperatures \cite{color1} and neutron-proton pairing in nuclear matter \cite{NuclMatt}, the question has become of particular importance and relevance by the recent discovery of graphene ~\cite{GeimScie,GeimNatu}, in which electrons at low energies behave like massless Dirac fermions with a pseudo-relativistic chiral property. One anticipates that the features of relativistic quantum physics to affect the magnetic ordering\cite{grapheneFerroTher} and the superconducting pairing \cite{grapheneSuperTher} in a fundamental way, and hence their mutual interplay and evolution. The aim of the present study is to address this question within a model based on graphene. We present zero temperature phase diagrams of relativistic FS in graphene with s-wave and p+ip-wave symmetry (see Fig.~\ref{fig1}c,d) which explore their fundamental difference from those of an ordinary FS. We show that FS in graphene provides the possibility for realizing a new gapless condensate phase in which Cooper pairing occurs between chiral carriers of opposite types, electron-like (n) and hole-like (p), which belong to different Dirac conduction and valance bands (see Fig.~\ref{fig1}a).

In the case of an ordinary s-wave superconductor, the presence of an spin-splitting field ~\cite{Chandra,Izuyama} causes a mismatching of the Fermi surfaces of the spin-down and spin-up electrons, and hence suppresses the number of paired states. This can lead to a first order quantum phase transition from superconducting (S) state to the normal (N) state in the so-called Clogston-Chandrasekhar limit specified by a critical field, $h_c =\Delta_{0}/\sqrt{2}$, with $\Delta_{0}$ being the S pairing gap at zero temperature and zero field. For the fields below $h_c$, FS exhibits properties of a homogeneous BCS condensate. There can also be existed homogenous S state with $\Delta_s<h$, known as Sarma or breached pairing state ~\cite{Chandra}, which involves pairing between the spin subband with a smaller Fermi surface and that with a larger Fermi sphere by curving out of a gap inside the larger Fermi sphere. This leaves the exterior part of the larger Fermi sphere unpaired, and give rise to the gapless property. Sarma state exhibits an instability, which can be avoided, for instance, by allowing a finite-range momentum dependence of attractive interaction between fermions in a single band Fermi system ~\cite{Wilczek}, and interband pairing in the case of two bands Fermi system ~\cite{Zhuang}. Here we demonstrate that in graphene-based FS, the new relativistic p-n superconducting phase dominates over Sarma state for s-wave symmetry, and over both BCS and Sarma states for p+ip-wave symmetry at low levels of doping. We find that, unlike the s-wave p-n phase, the corresponding p+ip condensate is stable and can be preserved in much strong exchange fields as compared to a conventional FS state.

To be more specific, we consider a graphene sheet in the presence of a spin-splitting exchange field $h$ and the superconducting pairing interactions with an on-site coupling constant $g_{0}$ and off-site nearest neighbor coupling constant $g_{1}$. In addition to the possibility of an intrinsic superconductivity, with plasmon or phonon mediated pairing interactions, in graphene ~\cite{grapheneSuperTher}, progress have already been made in  proximity-inducing superconductivity by fabrication of transparent contacts between a graphene monolayer and a superconductor (see for instance, Refs. \cite{grapheneSuperExper}). Similarly, ferromagnetic correlations can be induced by using an insulating ferromagnetic substrate or by F metals or added magnetic impurities on top of the graphene sheet ~\cite{grapheneFerroInduc}.
The induced magnetization is expected to be directed in-plane of graphene sheet, which allows for neglecting the orbital effects~\cite{orbitalfrustration}.

To describe the on-site and off-site FS of $\pi$ electrons in the honeycomb lattice structure of a monolayer graphene, we consider the following tight-binding Hamiltonian~\cite{grapheneSuperTher}
\begin{eqnarray}
H&=&-\sum_{i\sigma}(\mu+\sigma h)\hat{n}_{i\sigma}-t\sum_{\langle ij\rangle \sigma}(a_{i\sigma}^{\dagger}b_{j\sigma}+h.c.)\nonumber\\
&-&\frac{g_{0}}{2}\sum_{i\sigma}\left[a_{i\sigma}^{\dagger}a_{i\sigma}a_{i-\sigma}^{\dagger}a_{i-\sigma}+b_{i\sigma}^{\dagger}b_{i\sigma}b_{i-\sigma}^{\dagger}b_{i-\sigma}\right]\nonumber\\
&-&g_{1}\sum_{\langle ij\rangle}\sum_{\sigma,\sigma^{\prime}}a_{i\sigma}^{\dagger}a_{i\sigma}b_{j\sigma^{\prime}}^{\dagger}b_{j\sigma^{\prime}},
\label{HL}
\end{eqnarray}
where the operators $a_{i\sigma}/a^{\dagger}_{i\sigma}$ ($b_{i\sigma}/b^{\dagger}_{i\sigma}$) are on-site creation/annihilation operators for electrons of sublattice $A$ ($B$) with spin $\sigma=\pm$, respectively, $t\approx2.8$ eV is the hopping energy between nearest neighbor carbon atoms, $\hat{n}_{i\sigma}$ is the on-site particle density operator, and $\mu$ is the graphene chemical potential (we use the units such that $\hbar=1=k_{B}$). We constrain ourselves to the case of zero temperature, where the mean-field approximation in two dimensions is expected to have the highest validity~\cite{mermin}. Interestingly, we have noticed that the Hamiltonian (\ref{HL}) can equivalently describe BCS pairing of electrons from different layers of a double layer graphene, with levels of doping $\mu_1$ and $\mu_2$ in the layers 1 and 2 (see Fig.~\ref{fig1}b, lower panel). In this respect, the difference $\Delta \mu=(\mu_1-\mu_2)/2$ and the mean $(\mu_1+\mu_2)/2$ values in the double layer problem play the same rule as the exchange field $h$ and the chemical potential $\mu$ in FS, respectively. Thus, interlayer superconductivity in double layer graphene will exhibit the same physics as FS in monolayer graphene.

The on-site and the nearest neighbor parts of the interaction in (\ref{HL}) can be decoupled by the following spin singlet superconducting order parameters which have, respectively, s-wave and $p+ip$-wave symmetries
\begin{eqnarray}
\Delta_{s}=-g_{0}\langle a_{i\downarrow}a_{i\uparrow}\rangle=-g_{0}\langle b_{i\downarrow}b_{i\uparrow}\rangle,
\label{sw}
\end{eqnarray}
\begin{eqnarray}
\Delta_{p,ij}=-g_{1}\langle a_{i\downarrow}b_{j\uparrow}-a_{i\uparrow}b_{j\downarrow}\rangle.
\label{pw}
\end{eqnarray}
Replacing these mean field order parameters in $H$ and performing a Fourier transformation, we obtain the Hamiltonian in the space of 2D wave vector ${\bf k}$ of electrons. The excitation spectrum can then be obtained by a Bogoliubov diagonalization as
\begin{eqnarray}
E_{{\bf k \textit{l}}}^\sigma&=&\sqrt{(\xi_{{\bf k}}+\textit{l}\mu)^2+(\Delta_s+\textit{l}\xi_{{\bf k}}\Delta_p/t)^2}-\sigma h,
\label{exi}
\end{eqnarray}
where $\textit{l}=\pm1$ denote two branches of the spectrum and $\xi_{{\bf k+\bf K^{(\prime)}}}=v_F k$, in the linear approximation of the free electron spectrum around the two Dirac points $\bf K$ and $\bf K^{\prime}$; $v_F=\frac{3t}{2}$ is the Fermi velocity of Dirac particles, and $k$ is the magnitude of ${\bf k}$.
\par
\begin{figure}[!tb]
\begin{center}
\includegraphics[width=8.cm]{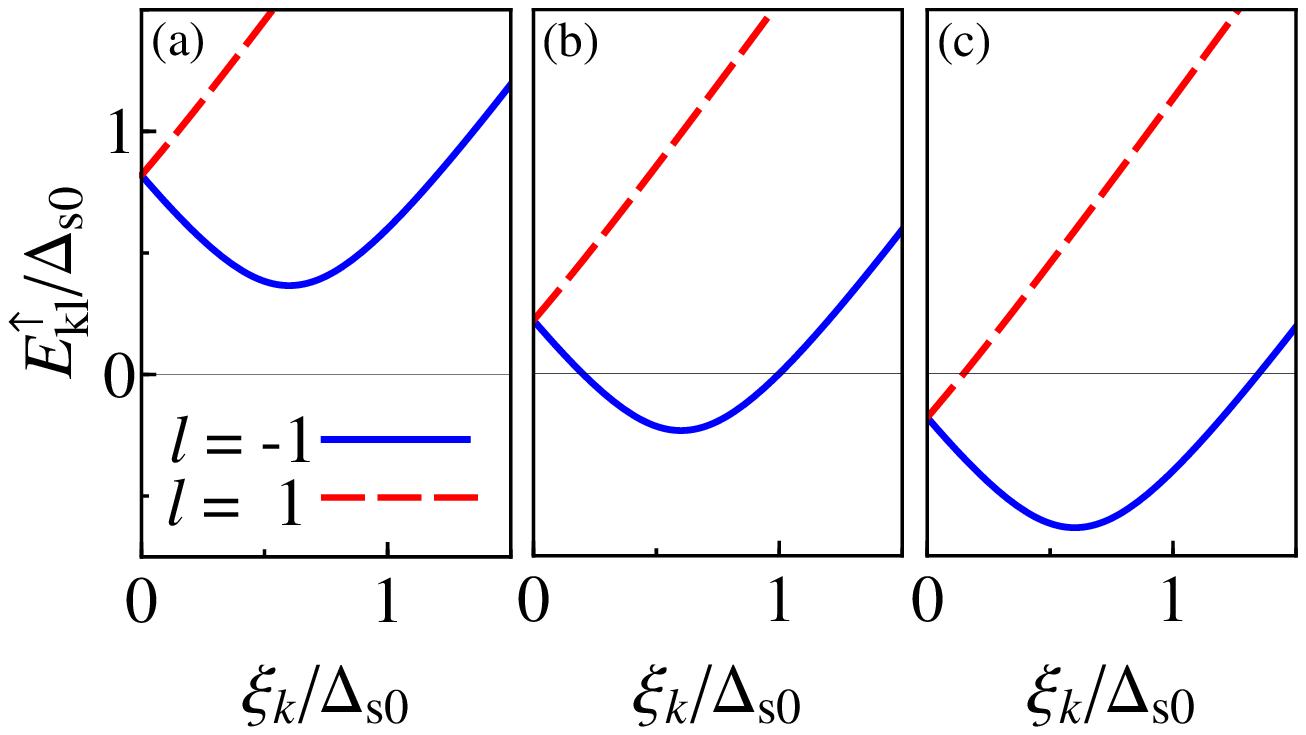}
\caption{(color online) The plot of quasi-particle excitation spectrum $E^{\uparrow}_{k \textit{l}}$ versus $\xi_k$ in the superconducting state. The dispersions correspond to the regimes of (a) BCS pairing with no effective Fermi surface for both branches $\textit{l}=\pm1$, (b) Sarma pairing with two effective Fermi surfaces for the branch $\textit{l}=-1$ but no Fermi surface   for the other branch $\textit{l}=1$, and (c) p-n pairing with single Fermi surface in each branch $\textit{l}=\pm1$. }
\label{fig2}
\end{center}
\end{figure}

From the excitation spectrum, we calculate the thermodynamic potential $\Omega_S$ using the relation
\begin{eqnarray}
\Omega_S=E_0+\sum_{\bf \textit{l},\sigma }\int{d^2{\bf k}\over(2\pi)^2}
E_{{\bf k \textit{l}}}^\sigma \bigg[
2\Theta(-E_{{\bf k \textit{l}}}^\sigma)-1\bigg],\label{Ome}
\end{eqnarray}
where $E_{0}=\Delta_{s}^{2}/g_{0}+3\Delta_{p}^{2}/g_{1}$ is the condensation energy and $\Theta(x)$ is Heaviside Theta function.
In the following we will see that the presence of $\Theta$ term in Eq.(\ref{Ome}), which depends on whether $E_{{\bf k \textit{l}}}^\sigma$ is positive or negative, has crucial effect on the type of the pairing resulting from the gap equation. The conditions that cause $E_{{\bf k \textit{l}}}^\sigma$ to be either positive or negative can be expressed in terms of the topology of the effective Fermi surfaces, as we investigate now. From Eqs.(\ref{exi}), one can see that for $h\geq0$, $\Delta_s>0$ and $\Delta_p>0$, and for any value of $\textit{l}$, and $\mu$ the spectrum $E^{\downarrow}_{k \textit{l}}$ is always positive. This situation is similar to that of a conventional superconductor in which exciting a paired state requires a finite energy. In contrast, the spectrum $E^{\uparrow}_{k \textit{l}}$ is not positive always. Depending on the involved parameters, it can undergo change of sign once for $\textit{l}=1$ and either once or twice for $\textit{l}=-1$. In Fig. \ref{fig2} the possible excitation spectrums are shown. By analyzing the roots of the equation $E^{\uparrow}_{k \textit{l}}=0$, we find three distinct regimes for $h$ with their own Fermi surface topologies.

In the range
\begin{eqnarray*}
h<\frac{|\Delta_s-\mu\Delta_p/t|}{\sqrt{1+(\Delta_p/t)^2}},
\end{eqnarray*}
there is no effective Fermi surface for both $\textit{l}=\pm1$, as can be seen in Fig. \ref{fig2}a. In this regime the pairing is of BCS type for both s-wave and p+ip-wave cases. For $\Delta_p=0$ and $\Delta_s\neq0$, this regime corresponds to an isotropic BCS superconductivity. Including a nonzero $\Delta_p$, can decrease slightly the area of this BCS phase in the phase diagram of s-wave FS.

The second regime occurs in the range
\begin{eqnarray*}
\frac{|\Delta_s-\mu\Delta_p/t|}{\sqrt{1+(\Delta_p/t)^2}}<h<\sqrt{\mu^2+\Delta^2_s},
\end{eqnarray*}
where the pairing is of breach-pairing type. For $\textit{l}=1$, there is no effective Fermi surface as in BCS regime. While for the branch $\textit{l}=-1$ there are two roots for $E^{\uparrow}_{k \textit{l}}=0$, and correspondingly two effective Fermi surfaces as is shown in Fig. \ref{fig2}b. This regime represents to Sarma phase in the phase diagram. For $\Delta_s=0$ and $\Delta_p\neq0$, as $\mu\rightarrow0$, the corresponding region in the phase diagram vanishes. Also for $\Delta_s\neq0$ and $\Delta_p=0$, this Sarma phase vanishes as $\mu\rightarrow0$ (see Fig.~ \ref{fig1}a,b).

The third range $h>\sqrt{\mu^2+\Delta^2_s}$ characterizes a new regime for which each of $\textit{l}=\pm1$ branches contains one effective Fermi surface (see Fig. \ref{fig2}c). Importantly, one can see that the superconductivity in this regime is resulted by pairing between normal Fermi surfaces of carriers of opposite spin, when they lie in different type, conduction or valence, subbands.
In this regime ferromagnetic graphene presents a spin chiral material, in which opposite-spin carriers are of different types, electron-like and hole-like\cite{grapheneFerroSuper}.

\begin{figure}[!tb]
\begin{center}
\includegraphics[width=8.cm]{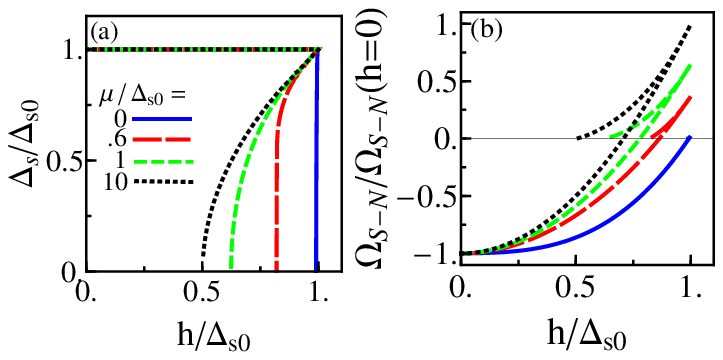}
\caption{ (color online) (a) Nonzero solutions of s-wave pairing gap equation versus exchange field $h$ for different levels of doping $\mu$. We have set $\Delta_{s0}=\Lambda/100$. (b) Difference in thermodynamic potentials of N and S states versus $h$ calculated at the same values of $\mu$ as in (a).
\label{fig3}}
\end{center}
\end{figure}
According to the above three ranges for exchange field, the gap equation has different solutions, which we analyze them in the following.
We obtain the gap equations by minimizing the thermodynamic potential Eq.(\ref{Ome}) with respect to the order parameters $\Delta_s$ and $\Delta_p$,
$\partial\Omega/\partial\Delta_{s,p}=0$.
In general, these are two coupled equations, from which the self-consistent solutions of the gaps can be calculated. We investigate the s-wave and p+ip-wave solutions separately, so in this case gap equations are not coupled.
Let us start with the case when only the on-site attractive interaction exists ($g_0\neq0$) and $g_1=0$. In this case, for s-wave condensate from Eqs.(\ref{Ome}) we obtain the gap equation as
\begin{equation}
\left(\frac{1}{g_0}-\sum_{\textit{l}}\int{d^2{\bf k}\over
(2\pi)^2}\frac{\Theta(E_{{\bf k \textit{l}}}^\uparrow)}{ E_{{\bf k \textit{l}}}^\uparrow+h}\right)\Delta_s=0.
\label{gap0}
\end{equation}
Similarly for the off-site nearest neighbor superconductivity ($g_0=0$, $g_1\neq0$), with $p+ip$-wave symmetry, we obtain
\begin{equation}
\left(\frac{t^2}{g_1}-\sum_{\textit{l}}\int{d^2{\bf k}\over
(2\pi)^2}\frac{\xi^2_{\bf k}}{3}\frac{\Theta(E_{{\bf k \textit{l}}}^\uparrow)}{E_{{\bf k \textit{l}}}^{\uparrow}+h}\right)\Delta_p=0.
\label{gap1}
\end{equation}
\begin{figure}[!tb]
\begin{center}
\includegraphics[width=8.cm]{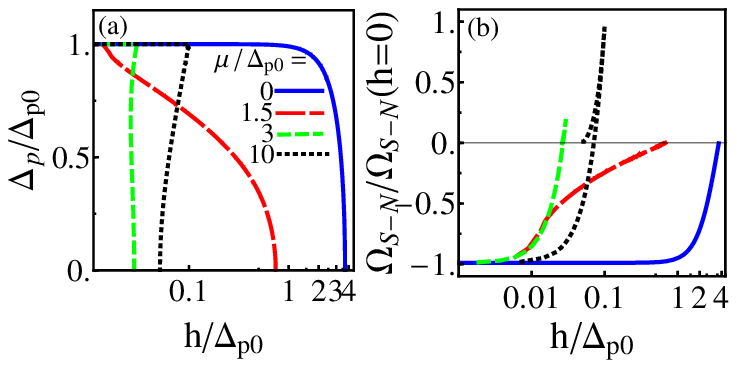}
\caption{(color online) The same as in Fig.~\ref{fig3}, but for p+ip-wave symmetry. Here we have set $\Delta_{p0}=\Lambda/100$.\label{fig4}}
\end{center}
\end{figure}
In Fig.~\ref{fig3}a, the nonzero solutions of Eq. (\ref{gap0}) as a function of $h$ are shown for different levels of doping. At high dopings $\mu\gg \Delta_{s0}$, the situation is similar to an ordinary s-wave FS wherein only BCS and Sarma solutions are possible. The BCS solutions extent over $0\leq h \leq \Delta_{s0}$ with a constant gap $\Delta_{s}=\Delta_{s0}$.  The Sarma solutions appear in the range $1/2\leq h/\Delta_{s0} \leq 1$, for which the pairing gap increases with increasing $h$. Lowering $\mu$ has no effect on BCS solutions, but leads to a suppression of Sarma states. At a critical doping with $\mu=\sqrt{h^2-\Delta^2_{s}}$ p-n pairing solutions also begin to appear within a sharply narrow range of the exchange field. Lowering $\mu$ further leads to a growing of p-n states and further suppressing of Sarma solutions. Approaching Dirac point $\mu=0$, p-n states dominates fully over Sarma states. We have analyzed the stability of these solutions in Fig.~\ref{fig3}b, by plotting the difference in thermodynamic potentials of N and S states, $\Omega_{S-N}$, as a function of $h$ for different values of $\mu$. We see that only BCS solutions are stable and that is up to the critical field $h_c$, wherein graphene FS undergoes a first order phase transition into N state. Interestingly, the range of stability of BCS states  increases with decreasing the doping and tends to be over the full range of $0\leq h\leq\Delta_{s0}$ in the limit of $\mu\ll \Delta_{s0}$.
We note that Sarma and p-n pairing states are always unstable for s-wave pairing. The resulting S-N phase diagram of this spin-singlet FS in the plane $\mu-h$ is presented in Fig.~\ref{fig1}c, in which the dashed line indicates the boundary between the stable (left side) and unstable (right side) BCS states.

The results for nonzero solutions of gap equation (\ref{gap1}) for p+ip-wave symmetry are presented in Fig.~\ref{fig4}a. At high $\mu$ the solutions have the same structure as in the s-wave case. They consist of Sarma states at high exchange fields and BCS states with a critical exchange field $h_c=(\mu/t) \Delta_{p0}/\sqrt{2}$. Decreasing $\mu$ causes a decrease in the range of not only Sarma states but also BCS states. At a critical doping Sarma states are disappeared completely, and at the same time p-n solutions start to emerge. From this point on the gap equation support single-valued solutions consisting of BCS and p-n solutions, at low and high exchange fields, respectively. Upon approaching Dirac point there remains only single-valued p-n solutions, which extends up to a large critical field $h_c=\Lambda(1 -9N_1/2\Lambda)^\frac{1}{3}$, where $N_1=2\pi v_F^2t^2/g_1$ and $\Lambda$ is the highest energy scale as the cutoff in the integration over the energy. As in the case of s-wave pairing, we have examined the stability of these solutions by plotting $\Omega_{S-N}$ as a function of $h$ for different values of $\mu$. The results presented in Fig.~\ref{fig4}b, show that in addition to BCS state whose range of stability (like the range of their existence) decreases with lowering $\mu$, p-n states are also stable for all values of $\mu$. We have presented the resulting phase diagram in Fig.~\ref{fig1}d . As in Fig.~\ref{fig1}c, we have depicted the boundary between the stable and unstable S states, whose dashed and solid parts indicate the first- and the second-order S-N phase transitions, respectively. We note that the above presented dependence of $\Delta_{s,p}$ versus $h$ can be used to distinguish between three different type of phases. Crossing a boundary between BCS and Sarma phases, the derivative of $\Delta_{s,p}$ versus $h$ will undergo a jump. While a p-n phase is distinguishable by its existence at exchange fields larger than the pairing gap, the fields for which a normal phase is expected in the ordinary FS.

In conclusion, we have developed a mean field theory of relativistic-like ferromagnetic superconductivity in graphene with an on-site and a nearest neighbor off-site pairing attraction. The resulted zero temperature phase diagrams for s-wave and p+ip-wave FS explore the existence of a new gapless homogeneous S phase, which is formed by pairing of chiral carriers of opposite types, electron-like and hole-like. We have found that the p-n pairing with p+ip-wave symmetry presents a stable condensate which, at low levels of doping, dominates the phase diagram up to exchange fields much larger than the superconducting pairing gap.

We acknowledge useful discussions with S. Abedinpour, C. W. J. Beenakker and A. G. Moghaddam. We also gratefully acknowledge support by the Institute for Advanced Studies in Basic Sciences (IASBS) Research Council under grant No. G2011IASBS110.

\end{document}